\documentclass[aps,prd,twocolumn]{revtex4}
\usepackage{epsfig}
\usepackage{latexsym,amsmath,amssymb,amstext}

\newcommand{\ze}{{\cal{Z}}}
\newcommand{\phat}{\hat{\phi}}
\newcommand{\mhat}{\hat{m}}
\newcommand{\wrt}{with respect to }

\begin{document}
\title{Thermodynamics and Finite size scaling in Scalar Field Theory}
\author{Debasish\ \surname{Banerjee}}
\email{debasish@theory.tifr.res.in}
\affiliation{Department of Theoretical Physics, Tata Institute of Fundamental
         Research,\\ Homi Bhabha Road, Mumbai 400005, India.}
\author{Saumen\ \surname{Datta}}
\email{saumen@theory.tifr.res.in}
\affiliation{Department of Theoretical Physics, Tata Institute of Fundamental
         Research,\\ Homi Bhabha Road, Mumbai 400005, India.}
\author{Sourendu\ \surname{Gupta}}
\email{sgupta@tifr.res.in}
\affiliation{Department of Theoretical Physics, Tata Institute of Fundamental
         Research,\\ Homi Bhabha Road, Mumbai 400005, India.}

\maketitle

\section{Introduction}
In this work we consider the 1-component real scalar $\phi^4$ theory in 4 space-time dimensions on the lattice and investigate the finite size scaling of thermodynamic quantities to study whether the thermodynamic limit is attained. 
The results are obtained for the symmetric phase of the theory.

\section{Thermodynamic set up}
The thermal expectation value of an operator ${\cal{O}}$ is
\begin{equation}
 \langle {\cal{O}(\varphi)} \rangle = \frac{1}{{\cal{Z}}(\beta)} \int {\cal D}\varphi e^{-\beta H}{\cal{O(\varphi)}}
\end{equation}
where ${\cal{Z}}(\beta)$ is the partition function. For our finite temperature study, we have kept the temporal side fixed at $T=\beta^{-1}=(N_t a_t)^{-1}$ and varied the spatial extent $L=N_s a_s$ by going to higher values of $L T$ to examine the thermodynamic limit. Energy density $E$ and pressure $P$ are obtained by the following derivatives of the partition function: $E=T^2/V\left. \partial \ln \ze/\partial T\right|_V=-1/(N_s^3 a_s^3 N_t) \left. \partial \ln \ze / \partial a_t \right|_V$ and $P=T \left. \partial \ln \ze/\partial V\right|_T=1/(3 N_s^3 a_s^2 N_t a_t) \left. \partial \ln \ze/\partial a_s\right|_T$ where $a_s$  and $a_t$ are the lattice spacings along the spatial and temporal directions.
\newline
The continuum Euclidean action for the scalar $\phi^4$ theory is
 \begin{equation}
\small{
  S[\phi]=\int dt \int d^3 x\left \lbrace\frac{1}{2}\partial_{\mu}\phi(x) \partial_{\mu} \phi(x)+\frac{1}{2}m_0^2 \phi(x)^2+\frac{g_0}{4!}\phi(x)^4 \right\rbrace}
\label{action}
  \end{equation}

The following redefinitions are made to trade off the set of couplings ($g_0,m_0$) for ($\kappa,\lambda$):
\begin{eqnarray}
\nonumber \kappa &=& \frac{1-2\lambda}{3\xi+\xi^3+((a_tm_0)^2/2) \xi^3};~~\kappa_t = \kappa_s \xi^2 = \kappa \xi^3;\\
\lambda &=& \frac{g_0}{24}\xi^3 \kappa^2;
\phat = a_t \frac{\phi}{\sqrt{\kappa}};~~\mhat_0 = a_t m_0
\label{bareparam}
 \end{eqnarray}
where $\xi=a_s/a_t$ is the anisotropy parameter. This yields the conventional form of the action for doing simulations:

\small{
\begin{eqnarray}
\label{latact}
 \nonumber S[\phat]&=&\sum_{x_{\alpha}} \Bigl\{ -\kappa_s \sum_{i=1}^3 \phat(x_{\alpha}+a_i)\phat(x_{\alpha})-\kappa_t \phat(x_{\alpha}+a_t)\phat(x_{\alpha}) \Bigr.\\
 && \qquad  +\Bigl. \phat(x_\alpha)^2 + \lambda(\phat(x_\alpha)^2-1)^2 \Bigr\} 
\end{eqnarray}
}
\normalsize{
 Since our simulations are on lattices with $\xi=1$, we calculate the relevant thermodynamic expressions for anisotropic lattices and then set the spacings to be equal. We renormalise the pressure to be zero at $T=0$ by subtracting the corresponding zero temperature operators from their finite temperature counterparts. This procedure is adopted to renormalise all the operators appearing in the other thermodynamic quantities as well and is denoted by adding the subscript ``subt'' to the operators. Finally, the expressions for the energy density $E$, pressure $P$ and $E-3P$ in this formulation is:
}
\small{
\begin{eqnarray}
\nonumber E a_t^4 &=& \frac{1}{\xi^3} \left\lbrace-\xi\left\langle\left.\frac{\partial S}{\partial \xi} \right|_{a_t}\right\rangle_{subt}+\left\langle \left. a_t \frac{\partial S}{\partial a_t} \right|_{\xi}\right\rangle_{subt} \right\rbrace\\
\nonumber   P a_t^4 &=& -\frac{1}{\xi^2}\left\langle\left.\frac{\partial S}{\partial \xi} \right|_{a_t}\right\rangle_{subt}\\
   (E-3P) a_t^4 &=& \frac{1}{\xi^3}\left\langle \left. a_t \frac{\partial S}{\partial a_t} \right|_{\xi}\right\rangle_{subt}
\label{tdycquant}
\end{eqnarray}}
\normalsize{\hspace{-1.5mm}where bracketed quantities denote thermal averages. Note that the quantities $E a_t^4$,~$P a_t^4$ and $(E-3P)a_t^4$ are all dimensionless quantities and will henceforth be referred to as $E$,~$P$ and $E-3P$ respectively in lattice units.~The derivatives of the couplings of the theory \wrt $\xi$ are the anisotropy coefficients.
\newline
The renormalisation conditions for the theory involve fixing the 2- and 4-point  renormalised vertex functions at zero momenta as follows:
\begin{equation}
 \mhat_R^2 = - \Gamma^{(2)}_R(0);~g_R = -\Gamma_R^{(4)}(0,0,0,0)
\end{equation}

\begin{figure}
\begin{center}
\includegraphics[scale=0.25]{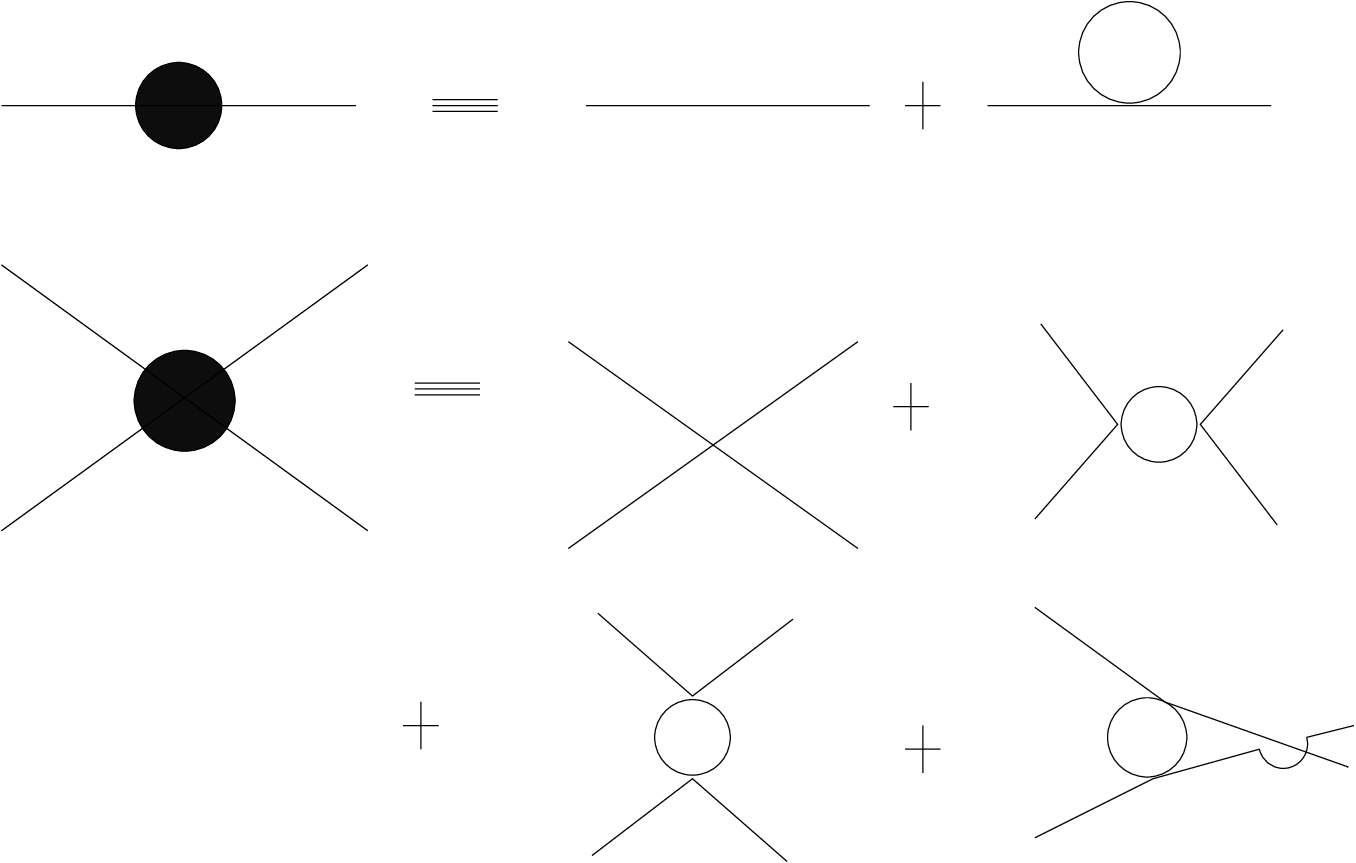}
\caption{The Feynman diagrams to 1-loop order that need to be calculated. The first set of diagrams is for the 2-point vertex function while the second one is for the 4-point vertex function.}
\label{fig:feyndiag}
\end{center}
\end{figure}

Equation (\ref{tdycquant}) requires evaluation of the following derivatives:
\begin{eqnarray}
 d_{\kappa_s}&=&\left. a_t \frac{\partial \kappa_s}{\partial a_t} \right|_{\xi};~d_{\kappa_t} = \left. a_t \frac{\partial \kappa_t}{\partial a_t} \right|_{\xi};~d_\lambda = \left. a_t \frac{\partial \lambda}{\partial a_t} \right|_{\xi}\\
c_{\kappa_s}&=& \left. \kappa_s^{\prime}\right|_{a_t};~~~~~c_{\kappa_t} =  \left. \kappa_t^{\prime}\right|_{a_t};~~~~~~c_\lambda =  \left. \lambda^{\prime}\right|_{a_t}
\end{eqnarray}
 where $f^\prime= \partial f/\partial \xi$. The first three derivatives can be completely specified using  $\beta=a_t \left. \partial g_0/\partial a_t \right|_{\xi}$ and  $\gamma=a_t \left.\partial \hat{m_0}^2/\partial a_t \right|_{\xi}$ functions. We have used the 1-loop expressions for the $\beta$ and $\gamma$ functions of the lattice theory available in literature \cite{MM}. The latter three involve the anisotropy coefficients $g_0^\prime=\left. \partial g_0/\partial \xi \right|_{a_t}$ and $(\hat{m_0}^2)^\prime=\left. \partial \hat{m_0}^2/\partial \xi \right|_{a_t}$, whose expressions do not exist in literature for this scalar theory. We have evaluated the anisotropy coefficients upto 1 loop order by summing the Feynman diagrams shown in fig \ref{fig:feyndiag} to be:

\begin{eqnarray}
\nonumber K(x) &=& \left[(3\mhat_R^2+12)I_0^4(2x)-12I_0^3(2x)I_1(2x)\right]\\
\nonumber  \left. (\mhat_0^2)^{\prime} \right|_{a_t}&=&\frac{g_R}{2}\int_0^{\infty}x dx e^{-(\mhat_R^2+8)x}K(x)\\
\left. g_0^{\prime} \right|_{a_t} &=& -\frac{3}{2}g_R^2 \int_0^{\infty} x^2 dx e^{-x(\mhat_R^2+8)}K(x)
\end{eqnarray}
where $I_n(x)$ denotes the modified Bessel function of order n. The calculation of these coefficients was done using Mathematica which could handle large range of numbers, much beyond the capacity of usual C compilers. This was specially important since the numerical values depend on a delicate cancellation among Bessel functions. The anisotropy coefficients evaluated at the simulation points is listed in table \ref{table:Params}.

\section{Numerical Methods}

\begin{table}
 \caption{Parameter values at each of the different points.~$g_R$ is calculated in 1 loop perturbation theory.~For comparison,~estimate of mass in 1 loop perturbation theory is also shown.}
 \label{table:Params}
 \begin{center}
 \begin{tabular}{|c|c|c|c|c|}
 \hline
 No.& A & B & C & D \\
\hline
$\kappa$ & $0.2516$ & $0.2526$ & $0.253452$ & $0.25438$\\
\hline
$\lambda$ & $0.011153$ & $0.0117$ & $0.0122$ & $0.0134$ \\
\hline
$g_R$ & $3.1772$ & $3.18955$& $3.16478$& $3.17108$ \\
\hline
$ma$ & $0.2853 $& $0.2376 $& $0.1897 $& $0.1416 $ \\
    & $\pm 0.0002$ & $\pm 0.0003$ & $\pm 0.0002$ & $\pm 0.0002$\\
\hline
$ma$ (in 1 loop) & $0.3012$ & $0.2586$ & $0.2172$ & $0.1834$\\
\hline
$g_0^\prime$ & $-1.9372$ & $-2.06348$ & $-2.16451$ & $-2.34449$\\
\hline
$(\hat{m_0}^2)^\prime$ & $0.367901$ & $0.369738$ & $0.367193$ & $0.368173$\\
 \hline
 \end{tabular}
 \end{center}
 \end{table}

A overrelaxation+Metropolis (OR+M) algorithm was used to generate the equilibrium configurations on which the operators were measured. The OR+M algorithm caused the correlation time at critical point to reduce by a factor of 10 compared to the Metropolis and the time to generate independent configurations was $0.4$s. Configurations separated by $3 \tau_{int}$ were considered to be independent, where $\tau_{int}$ was the Monte-Carlo time for the autocorrelation function to decay to $1/e$.
To check the accuracy of our program,~simulations were done at ($\lambda=0.275376,\kappa=0.288998$) where we measured the vacuum expectation value of the $\phi$ field and its susceptibility,
and compared it with the results obtained in \cite{balog}.



The jackknife method was used to calculate the errors on the operators. It was checked that the block size used was much above the minimum bin size over which the variance showed a plateau.

\begin{figure}
\begin{center}
\includegraphics[scale=0.5]{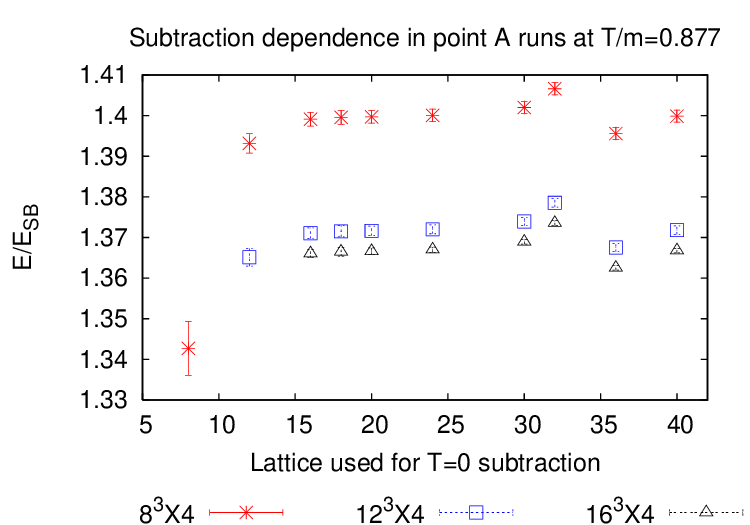}
\includegraphics[scale=0.5]{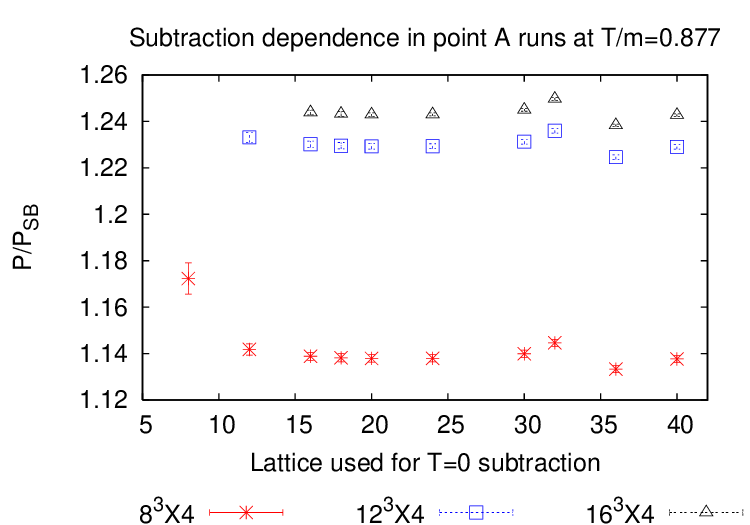}
\includegraphics[scale=0.5]{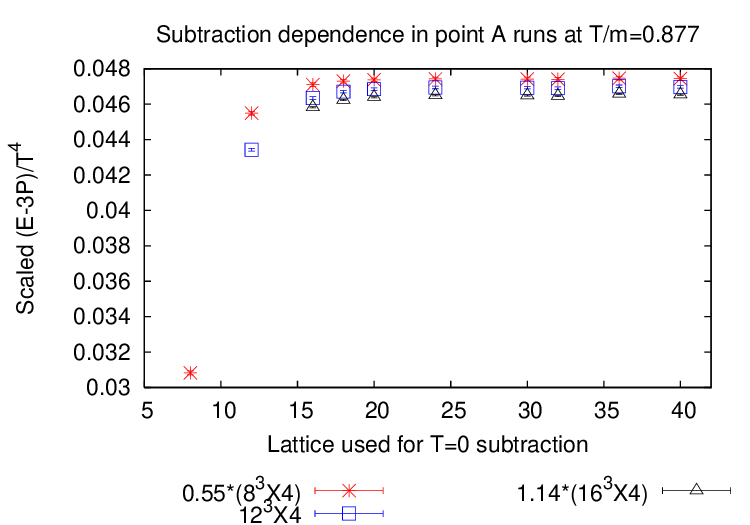}
\caption{Comparison of different subtraction schemes on the $E/E_{SB}$, $P/P_{SB}$ and $(E-3P)/T^4$ at point A}
\label{fig:subt}
\end{center}
\end{figure}

\section{Physics Results}
The simulations were done on Renormalisation Group (RG) trajectories of constant $g_R$. A point was chosen ($\kappa=0.2516,\lambda=0.011153$) where the scaling violations could be neglected \cite{lw1}. The physical mass was extracted non-perturbatively at this point A. Points B, C and D with cut-off effects $4/5$,~$2/3$ and $1/2$ of point A and lying on the same RG trajectory were first constructed within 1-loop perturbation theory and then non-perturbatively tuned to get the desired $ma$. The parameter details of the points are given in table \ref{table:Params}. 

\begin{figure}
 \begin{center}
 \includegraphics[scale=0.5]{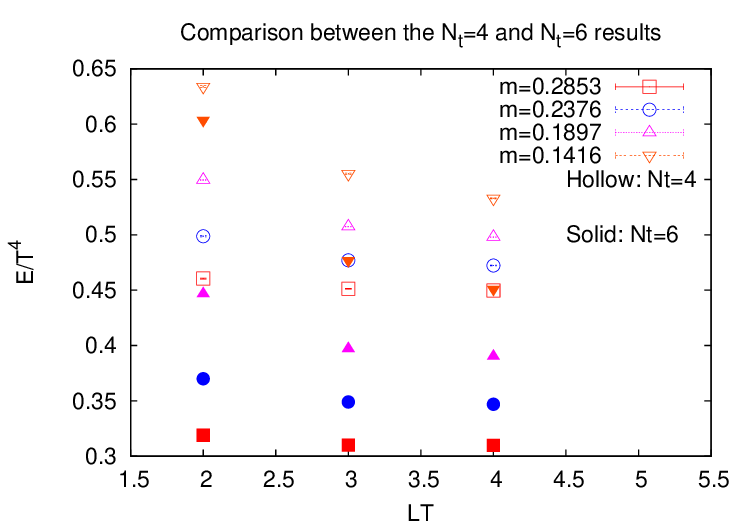}
 \includegraphics[scale=0.5]{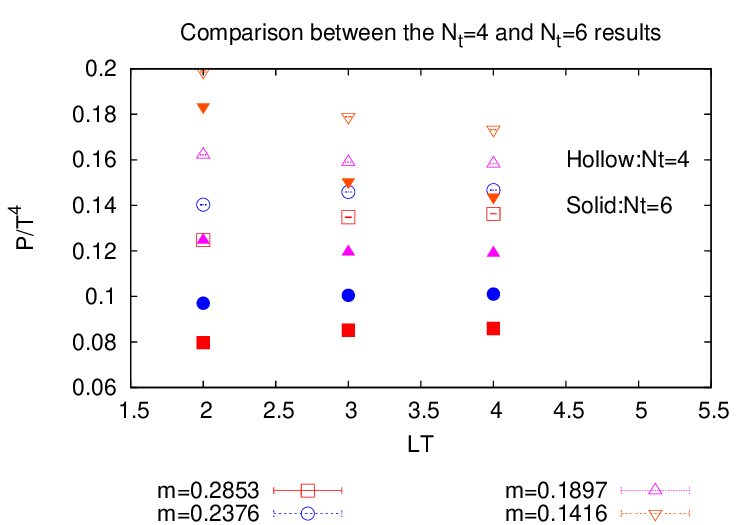}
 \includegraphics[scale=0.5]{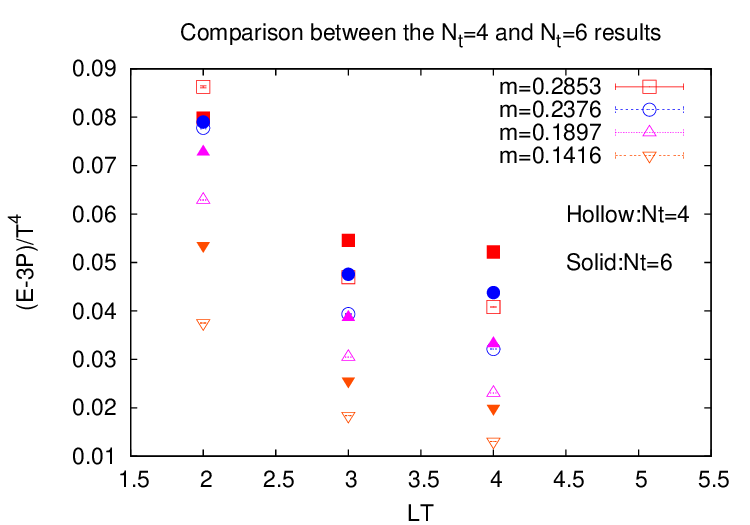}
 \caption{Scaling of the $E/T^4$, $P/T^4$ and $(E-3P)/T^4$ for $N_t=4$ and $N_t=6$.}
 \label{fig:scaling}
 \end{center}
 \end{figure} 

To estimate zero T subtraction, operators evaluated on lattices ranging from $40^4$ till $N_s^4$ were subtracted from the corresponding operators evaluated on the $N_s^3\times 4$ lattice at point A. The result is shown in fig \ref{fig:subt}. Our results indicate that a plateau is reached for all the thermodynamic quantities when the subtraction lattices used $N^4$ are much greater than $N_s^4$ except for the runs at $32^4$ and $36^4$ which seem to give erratic results. We checked that the problem was not in the analysis. In what follows, we do not include these two sets in the discussion of the thermodynamic limit. The result of subtracting with a $40^4$ lattice agrees with the results obtained by subtraction from the $N^4$ lattices except when $N$ is comparable to $N_s$.~Thus,~we chose to subtract using the largest lattices that we have simulated which is $40^4$.
\newline
The $E$, $P$ and $E-3P$  are plotted vs $LT$ in fig \ref{fig:scaling}. To establish the thermodynamic limit at constant physics we compare the $LT$ dependence of the $N_t=4$ and the $N_t=6$ lattices at each of the four different points. Our results show that as we decrease $ma$ at fixed $LT$ the finite size effects become larger,~as expected.~At fixed $m$ and for the same $N_t$,~the absolute value of the slope decreases with larger $LT$.~In particular,~for the $N_t=6$ lattices,~for energy density and pressure,~the local slope between the $LT=3$ and $LT=4$ is zero within error.~The $N_t=4$ and $N_t=6$ lattices at fixed $m$ can be compared by looking at the slopes of the thermodynamic quantities as a function of $LT$.~The local slope for $N_t=6$ would be lesser than the $N_t=4$ counterparts for the same range of $LT$.~This comparison also yields the expected results except in the case of the smallest mass.
\newline
For $E$ and $P$, the thermodynamic limit is reached for the $LT=4$ runs for the masses $ma=0.2853$ and $ma=0.2376$, since the local slope is zero within error. Even though this is not the case for $E-3P$, it certainly is close to the thermodynamic limit since the values approach a plateau. For example, at point A, the $E-3P$ for $N_t=6$ lattice decreases from $(6.16 \pm 0.02)\times 10^{-5}$ at $LT=2$ to $(4.212 \pm 0.006)\times10^{-5}$ at $LT=3$ and to $(4.025 \pm 0.004) \times 10^{-5}$ at $LT=4$. ~For the last two points where the finite size effects are large the situation is less clear. In terms of $m L$, our results show that $m L=4.55$ is not enough for the thermodynamic limit, while by $m L=5.70$ the limit is attained. We conclude that to reach the thermodynamic limit, one needs $LT \sim 4$ and $m L > 4.5$. We also found that the $T=0$ subtraction should be made with lattices with $N$ sufficiently greater than $N_s$. In this work we have not tried to take the continuum limit.
}

\end{document}